\documentstyle[prb,aps,eqsecnum,epsf]{revtex}
\begin{document}
\draft
\def \beq{\begin{equation}}
\def \eeq{\end{equation}}
\def \beqarr{\begin{eqnarray}}
\def \eeqarr{\end{eqnarray}}

\title{Dynamics of the Compact, Ferromagnetic $\nu=1$ Edge}

\author{A. Karlhede$^1$, K. Lejnell$^1$ and S. L. Sondhi$^2$.}

\address{
$^1$Department of Physics,
Stockholm University,
Box 6730, S-11385 Stockholm,
Sweden
}

\address{
$^2$Department of Physics,
Princeton University,
Princeton, NJ 08544, USA
}

\date{\today}
\maketitle
\begin{abstract}
We consider the edge dynamics of a compact, fully spin polarized state at filling factor
$\nu=1$. We show that there are two sets of collective excitations localized near
the edge: the much studied, gapless, edge magnetoplasmon but also an additional edge
spin wave that splits off below the bulk spin wave continuum. We show that both
of these excitations can soften at finite wave-vectors as the potential confining 
the system is softened, thereby leading to edge reconstruction by spin texture or 
charge density wave formation. We note that a commonly employed model of the
edge confining potential is non-generic in that it systematically underestimates
the texturing instability.
\end{abstract}

\section{Introduction}

Following the seminal work of Halperin \cite{halperin} and especially of Wen 
\cite{wen}, there has been a 
great deal of work building on the insight that there must be interesting low 
energy dynamics at the edges of incompressible quantum Hall systems. Much of
this work has focused on the presence of Luttinger liquid correlations at the
edges of model quantum Hall systems, leading to predictions that have considerable
support in experiments albeit with the perplexing feature that the latter seem
insensitive to the presence or absence of incompressibility in the bulk \cite{chang}.

In this paper we wish to revisit one of the simplest quantum Hall (QH) systems, that
at filling factor $\nu=1$. {\it Prima facie} this would seem to be an uninspiring
choice as the edge ground state correlations are known to be those of a Fermi
liquid and hence no interesting power laws are expected by theory, or seen in
the experiments thus far. Our motivation however is different and stems from the
identification of $\nu=1$ as the archetypical QH ferromagnet with non-trivial
bulk dynamics characterized by a density-topological density relation \cite{sondhi1}.
At issue here is whether this novel dynamics has a counterpart at the edges of
the system. We have previously shown in a Hartree-Fock/effective-action study
\cite{karlhede1} that this physics allows for a reconstruction of the $\nu=1$ edge
in which charge is moved outwards by texture formation as the confining potential
is softened--the same phenomenon takes place also in quantum dots \cite{tejedor}.
Here we will ask the logically prior question of how the low lying
but not necessarily gapless, edge excitations inclusive of spin are affected in 
the phase in which the edge is still sharp. We note that many of these results 
were noted previously in  conference contributions \cite{conferences} and in 
the interim there have appeared two publications \cite{brey,oaknin} with some 
overlapping content which we discuss in the main text. Edge reconstruction 
has also been studied within the Chern-Simon-Ginzburg-Landau theory 
\cite{leinaas} and more work on the reconstruction in quantum dots \cite{dots} 
has appeared---the latter is not directly connected to our results and we will 
not discuss it further. Finally, Milovanovic \cite{milovanovic} has attempted
to deduce the edge physics of ferromagnetic quantum Hall states from general
consistency principles applied to the bulk low energy effective action, but
she finds a mode structure that is at odds with our, microscopic, results. Currently,
we do not understand how her approach can be modified to remove this discrepancy.

The outline of the paper is as follows. In Section II we define the problem more
precisely and write down a set of generalized random phase approximation or
time dependent Hartree-Fock equations (TDHF) for the particle-hole pair dynamics. 
Next we check that in the absence of an edge these equations recover the well
known results \cite{bychkov,kh} for spin-waves in the bulk. In section IV we show
that for the problem with an ``ideal'' edge (to be defined below) the TDHF equations
give rise to a novel non-chiral edge spin wave (ESW) in addition to the much studied 
chiral edge magnetoplasmon (EMP).
In Section V we discuss these modes away from the ``ideal'' case. We find that they
can go soft as the confining potential at the edge is weakened and lead, naturally,
to a charge density wave reconstruction of the edge as well as the spin textured
reconstruction we have considered previously. We
also show that a commonly employed model for studying edge reconstruction is
non-generic and artificially suppresses textured reconstruction. In Section VI
we discuss in more detail how the ESW can be fitted into a continuum description
via a novel boundary condition. In Section VII we comment on the issues involved 
in going beyond the TDHF approximation, which turns out to be essentially exact
at long wavelengths, and then conclude with a brief summary and discussion 
of the experimental relevance of these results. 

\section{Hamiltonian and TDHF Equations}

We are interested in the problem of the $\nu=1$ QH state with a boundary where the
filling factor decreases to zero. To this end we consider a semi-infinite 2DEG
with a single boundary (Figure 1a). To model the confinement necessary to produce 
such an edge we assume the existence of a background compensating charge {\it in}
the plane which falls from the density of the bulk $\nu=1$ state to zero. 
\begin{figure}[htbp]
  \begin{center}
    \leavevmode
    \epsfxsize=8cm
    \epsfbox{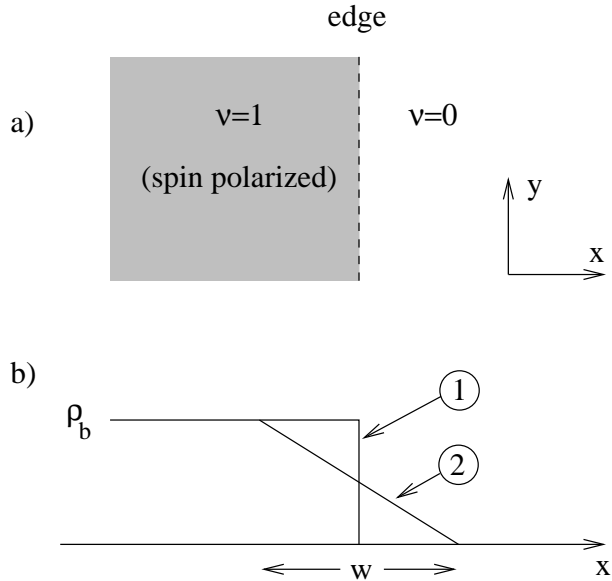}
  \end{center}
\caption{(a) Geometry of studied quantum Hall system; (b) Examples of in-plane 
compensating background charges that confine the electrons: (1) the ideal edge  
and (2) a softer confinement.}
\label{fig1}
\end{figure}

The particular choice denoted 1 in Figure 1b, defined more precisely in (\ref{compact})
below, has the effect of {\it exactly} canceling the Hartree potential of the
$\nu=1$ state with a sharp edge at $x=0$ which is thus favored.
We shall refer to the edge {\it with this choice of confining potential}
as the ``ideal edge''. More generally we shall consider background densities that
decrease more gradually but integrate to the same amount of charge as in the 
ideal edge,
e.g. the piecewise linear background sketched as 2 (in fact we shall see that this
choice is non-generic), as well as confining potentials independent of $y$
that arise from charges not in the plane of the electrons; the latter will turn out to be essentially different in
one important respect. These choices of confinement will tend to expand the area of 
the system and will compete with the tendency of the exchange to keep the electronic
state spin polarized and compact.
This softening of the confinement will eventually produce a reconstruction of the 
edge, but we will be interested here in the parameters for which the edge remains
compact.

To formalize this we will restrict the Hilbert space to the lowest Landau level of either
spin and choose Landau gauge ${\bf A} = Bx\hat {\bf y}$ in which the single particle states
are (setting the magnetic length $\ell=\sqrt{\hbar c/eB}=1$)
\beq
u_k(x,y) = {1\over \sqrt{\pi^{1/2} L}} e^{iky} e^{-(x-k)^2/2}
\eeq
with $e^{ikL}=1$ for a system of length $L$ made periodic along the edge. Expanding
the lowest Landau level electron field operator in this basis,
\beq
\Psi_\sigma (x,y) = \sum_k u_k(x,y) c_{k \sigma}
\eeq
( $\{c_k^\dagger, c_{k'}\}=\delta_{k,k'}$) we can write the many-body Hamiltonian
as
\beqarr
\hat{H} &=& {1 \over 2} \sum_{kpq \sigma \sigma'} V(p,q) c^\dagger_{k+q,\sigma}
c^\dagger_{k+p-q,\sigma'} c_{k+p,\sigma'} c_{k,\sigma} \nonumber \\
      && - \sum_{kp\sigma} V(p,0) \rho_b(k+p) c^\dagger_{k,\sigma} c_{k,\sigma}
\nonumber \\
      && + {1 \over 2} \sum_{kp} V(p,0) \rho_b(k) \rho_b(k+p) \nonumber \\
      && - {1 \over 2} g \mu_B B \sum_{k} ( c_{k \uparrow}^\dagger 
           c_{k \uparrow} - c_{k \downarrow}^\dagger c_{k \downarrow} ) \ .
\eeqarr
Here $V(p,q)=\tilde{V}(q,p-q,p,0)$ with $\tilde{V}(k_1,k_2,k_3,k_4)$ defined as
$\int d^2r d^2r' U({\bf r-r'}) \bar{u}_{k_1}({\bf r}) \bar{u}_{k_2}({\bf r'}) 
$\\$ u_{k_3}({\bf r'}) u_{k_4}({\bf r})$,
$\rho_b(k)$ is the occupation of the ``background orbital'' and defines the real
space background charge density via $\rho_b(x) = \int {dk\over2\pi} \rho_b(k)
{e^{-(x-k)^2}\over \sqrt{\pi}}$. The second term is the interaction of the electrons with
the background charge,  the third term is the c-number self interaction of the 
background charge density and in the fourth we will take the g-factor to be
positive. We will take the interparticle potential to be 
of the unscreened Coulomb form $U({\bf r}) = {e^2\over  |{\bf r}|}$ \cite{fn-eps},
which leads to the explicit expression,
\begin{equation}
V(p,q) = {e^2 \over L} \sqrt{2 \over \pi} e^{-[{q^2 \over 2} + {(p-q)^2
\over 2}]} \int_{-\infty}^{\infty} dy \, K_0(|qy|) e^{-(y^2 + 2(p-q)y)/2}
\end{equation}
for the matrix elements of interest.

The ideal edge is now defined by taking
\begin{equation}
\rho_b(k) = \Theta(-k)
\label{compact}
\end{equation}
which then exactly cancels the Hartree potential of the compact state
\beq
|G\rangle = \Pi_{k \le 0} c^\dagger_{k\uparrow} |0\rangle \ .
\eeq
This is an exact eigenstate of $\hat{H}$ for any $y$-invariant confining potential
by virtue of minimizing
the momentum along the $y$-direction. For the ideal edge Hamiltonian, this is the
ground state and in this paper we will be interested only in those choices of
confinement for which this continues to hold.

A first pass at the excitation spectrum is obtained by the Hartree-Fock (HF)  decoupling,
$\langle c^\dagger_{k\uparrow} c_{k\uparrow} \rangle = f_{\uparrow}(k)  = 
\Theta(-k)$ which yields the eigenvalues,
\beqarr
\epsilon_{HF}(k\sigma) &=& \Sigma_{p} V(p,0) [f_\uparrow(k+p) - \rho_b(k+p)]
\ \ \ {\rm Hartree} \nonumber \\
&& - \delta_{\sigma \uparrow} \Sigma_{p} V(k-p,k-p) f_\uparrow(p)
\ \ \ {\rm Exchange} \ .
\eeqarr
These are sketched in Figure 2 for the ideal edge and for a linear background
with (arbitrarily chosen) scale $w= 6.77$. The latter exhibits the beginnings
of a minimum outside the region occupied by the electrons, which leads to
a polarized reconstruction of the edge as discussed in the pioneering papers
\cite{macdonald1,chamonwen}.
\begin{figure}[htbp]
  \begin{center}
    \leavevmode
    \epsfxsize=8cm
    \epsfbox{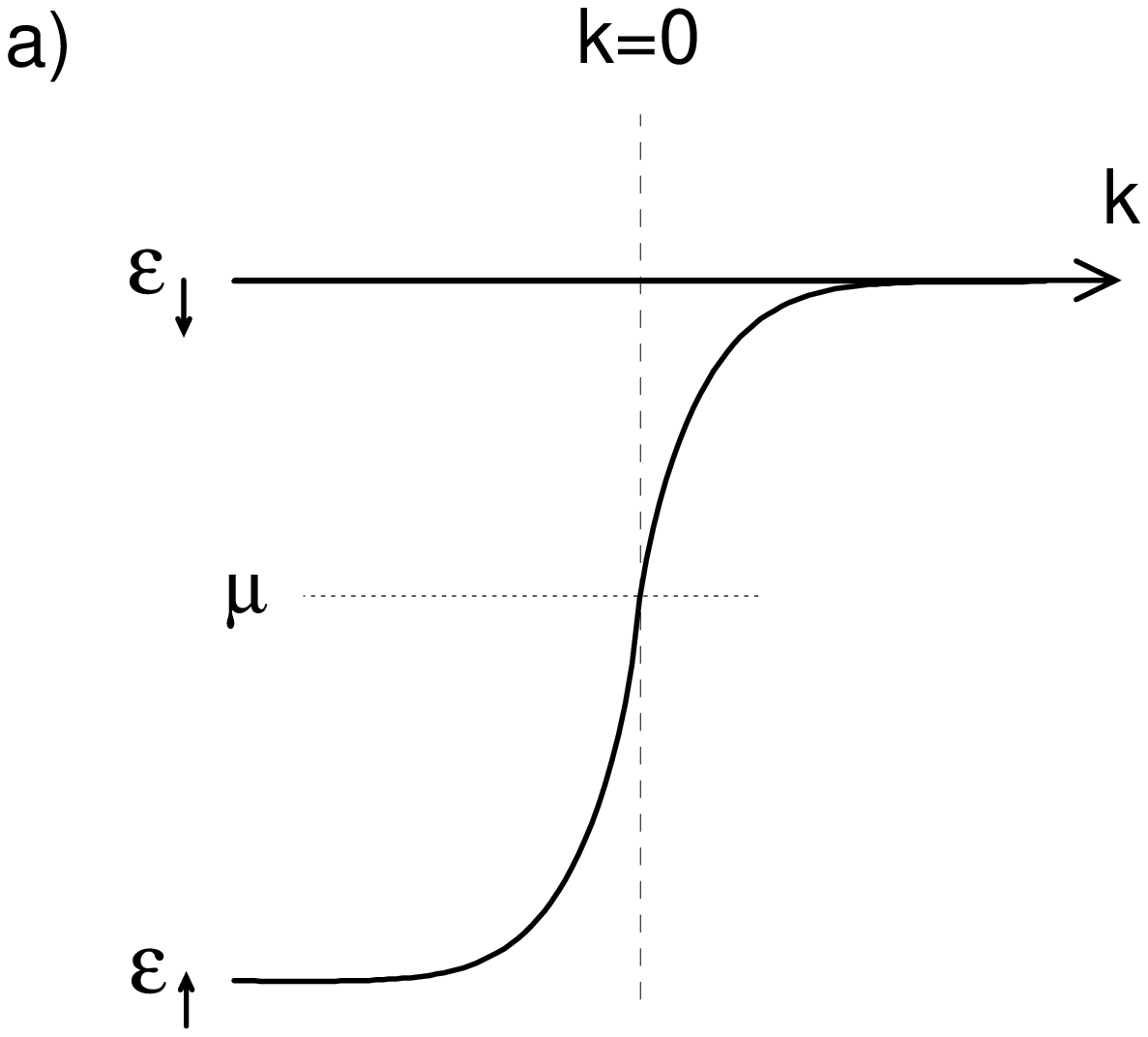}
    \epsfxsize=8cm
    \epsfbox{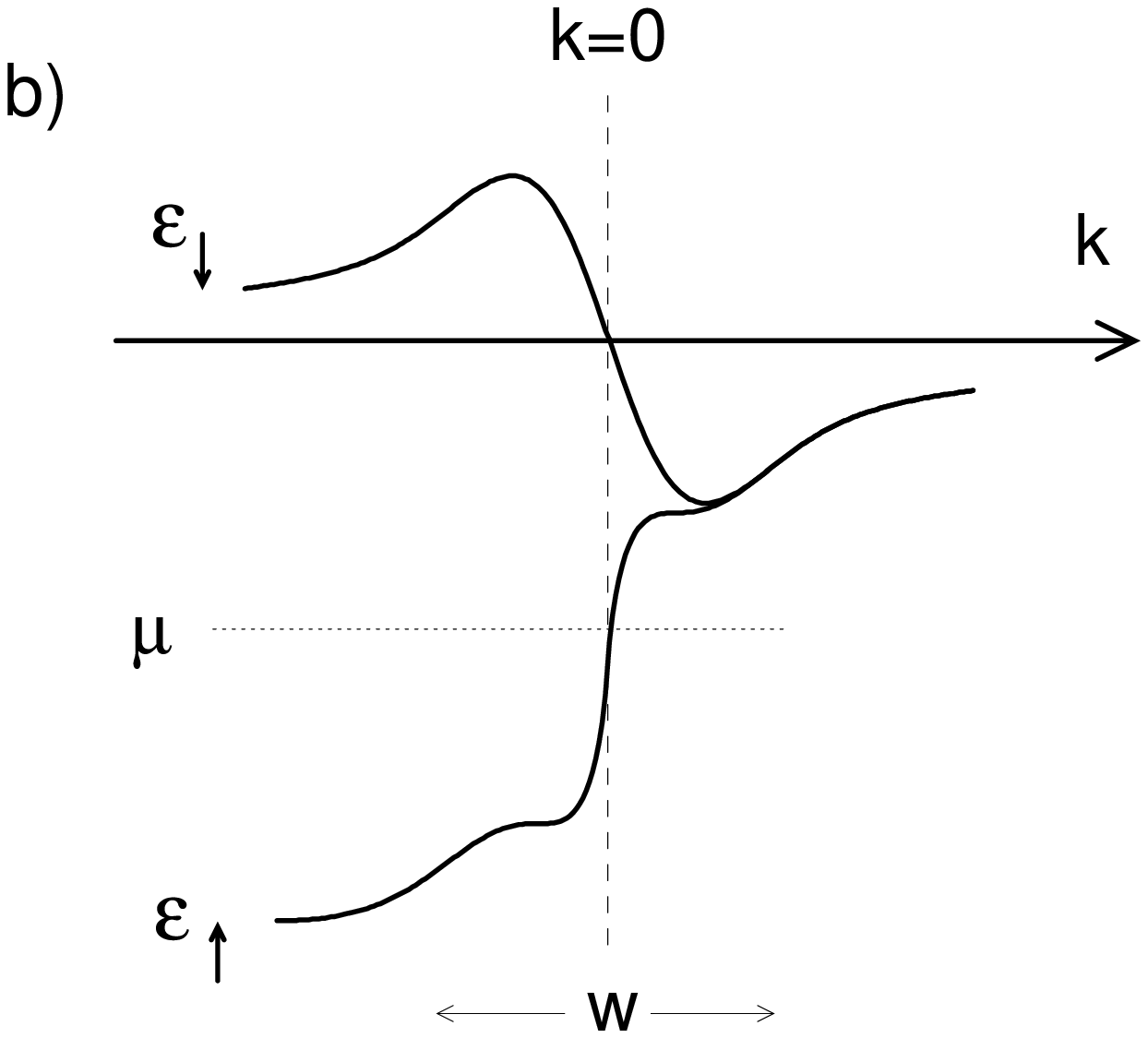}
  \end{center}
\caption{Hartree-Fock eigenvalues $\epsilon_{\sigma}(k) \equiv \epsilon_{HF}(k\sigma)$ for 
the ideal edge (a) and for a linear background with $w=6.77$. 
($\mu$ is the chemical potential.)}
\label{fig2}
\end{figure}

A second pass at the (neutral) excitation spectrum requires that we go beyond
HF to the time dependent Hartree-Fock approximation (TDHF) or the generalized 
random phase approximation. This consists of considering states which
involve one particle-hole excitation of the Hartree-Fock ground states and 
diagonalizing the full Hamiltonian in this restricted subspace. A convenient
way of deriving the TDHF equations is by the equations of motion method discussed,
for example, in \cite{pn}. In our case it yields, for the operators
$b^\dagger_{\sigma}(k,q) \equiv c^\dagger_{k+q \sigma} c_{k \uparrow}$, the
equation,
\beqarr
-i \partial_t b^\dagger_{\sigma}(k,q) &=& [\epsilon_{HF}(k+q \sigma)
- \epsilon_{HF} (k \uparrow)] b^\dagger_{\sigma}(k,q) 
+ \sum_{k'} [f_{\sigma}(k+q) - f_\uparrow(k)] \nonumber \\& & \times
[V(k-k'-q,k-k') - \delta_{\sigma \uparrow} V(k-k'+q,q)] b^\dagger_{\sigma}(k',q)
\ .
\label{tdhf}
\eeqarr
These describe the HF time evolution of the particle-hole states (first term) as well
as the scattering between them (second term). At this level of approximation, there
is no mixing between states with spin flips ($\sigma =\downarrow, \, 
\Delta S^z = -1$) and those without ($\sigma = \uparrow, \, \Delta S^z = 0$).

In the following, we will find it very useful, on account of the one dimensional nature
of the Landau gauge description, to think of Eqn (\ref{tdhf}) as describing the 
quantum mechanics
of a particle representing the particle hole pair that is located at $k$, carries 
quantum numbers $q$ and $\sigma$, hops between different spatial
locations with an effective kinetic energy $T_{\sigma q} (k-k')
=  L [f_{\sigma}(k+q) - f_\uparrow(k)][V(k-k'-q,k-k') - 
\delta_{\sigma \uparrow} V(k-k'+q,q)]$ and
experiences a potential $V_{\sigma q}(k)= \epsilon_{HF}(k+q \sigma)
- \epsilon_{HF} (k \uparrow)$. In the limit $L \rightarrow \infty$,
the Schr\"odinger equation for its motion becomes the integral equation, 
\beqarr
\int_{-\infty}^{0} {dk' \over 2 \pi}\,  T_{\sigma q} (k-k') 
\psi_{\sigma q}^{(\alpha)} (k') + V_{\sigma q}(k) 
\psi^{(\alpha)}_{\sigma q}(k)
= \epsilon^{(\alpha)}_{\sigma q} \psi^{(\alpha)}_{\sigma q}(k)
\ .
\label{schrodinger}
\eeqarr
Its solutions then define the TDHF eigenstates through their action on the HF 
ground state $|G\rangle$,
\begin{equation}
|\alpha; \sigma q \rangle = \int_{-\infty}^{0} {dk \over 2 \pi}\,
\psi^{(\alpha)}_{\sigma q}(k) b^\dagger_{\sigma}(k,q)
| G \rangle \ .
\end{equation}

\section{Bulk Modes}

In the absence of edges, there are no polarized $\Delta S^z = 0$ particle-hole
pair states. In Eqn (\ref{tdhf}) this is reflected in the trivial 
time evolution
when we put $\sigma = \uparrow$ and $f_\uparrow(k)=1 \, \forall k$. The
non-trivial dynamics is in the $\Delta S^z = -1$ sector, where $\sigma=\downarrow$. 
In this sector we need to solve (suppressing the spin
index),
\beqarr
\int_{-\infty}^{+\infty} {dk' \over 2\pi}\, T_q(k-k') 
\psi^{(\alpha)}_q(k') + V_q(k) \psi^{(\alpha)}_q(k)
= \epsilon^{(\alpha)}_q \psi^{(\alpha)}_q(k)
\ ,
\eeqarr
with $T_q(k-k') = -L V(k-k'-q,k-k')$ and $V_q(k) = 
\Delta_Z +  \int_{-\infty}^{+\infty} {dp \over 2 \pi}  \, L \, V(k-p,k-p)$ 
$\equiv \Delta_Z + \Delta_{\rm exch}$ {\it
independent} of $k$ and $q$, is the ``exchange enhanced spin gap''. (We have 
included the bare Zeeman gap, $\Delta_Z$, in $V_q(k)$.)

This is a translationally invariant problem with plane wave eigenstates,
$\psi^{(\alpha)} \propto e^{i \alpha k}$, $-\infty < \alpha < +\infty$ in
all $q$ sectors, whose energies are given by,
\begin{eqnarray}
\epsilon_q^{(\alpha)} &=& \Delta_Z + \Delta_{\rm exch} + M(\alpha,q) \nonumber \\
{\rm with} \ M(\alpha,q) &=&  -  \int_{-\infty}^{+\infty} {dk \over 2 \pi} \,
e^{i \alpha k} L \, V(k-q,k) \ .
\label{bulksw}
\end{eqnarray}

In our construction of these $\Delta S^z = 1$ states we have used orbitals with well
defined $y-$momentum and so it is clear that $\epsilon_q^{(\alpha)}$ is the
energy of a state with $y-$momentum $q$. The interpretation of $\alpha$ is a bit
obscure in this derivation---it appears as the momentum ``conjugate'' to the index of
the Landau gauge orbitals. However, one can check that $M(\alpha,q)$ is isotropic in
the  $(\alpha,q)$ plane, leading to the identification of $\alpha$ as the $x-$momentum
of the spin wave state. The resulting dispersion relation is identical with the result
obtained in \cite{bychkov,kh}. Note that $S^z$ is a good quantum number and
that the single spin flip states exhaust the $\Delta S^z=-1$ states in the
lowest Landau level. Hence the spin wave states that we have obtained by
diagonalizing the Hamiltonian in the basis of the single spin flip states
are exact eigenstates in the lowest Landau level.

\section{Modes of the Ideal Edge }

Returning to the problem of the ideal edge, we note that it is now possible to 
have
single particle-hole excitations with $\Delta S^z=0$, their wavefunctions 
$\psi_{\uparrow q}(k)$ having
support on $-q \le k \le 0$. The second new feature is that the $\Delta S^z=-1$ sector
no longer consists solely of single spin flip states, it being possible to excite
additional particle-hole pairs within the spin up Landau level. Consequently  our
analysis will now be approximate for both sectors. We should note that the qualitative
features of several of our results have also appeared in the numerical solution
of the TDHF equations by Franco and Brey \cite{brey}.

The  $\Delta S^z=0$ excitation, the edge magnetoplasmon (EMP), is well known and has been
much studied. Here we will derive its long wavelength dispersion in our particular
choice of confining potential. To this end we use the asymptotic forms,
\begin{eqnarray}
{1 \over e^2} T_{\uparrow q} (k-k') &=& - 
\log ({q \over k - k'})^2 \nonumber \\
{1 \over e^2} V_{\uparrow q} (k) &=& {k+q \over 2 \pi} \log({C\over (k+q)^2})
- {k \over 2 \pi} \log({C\over k^2})
\end{eqnarray}
($C=8 e^{2-\gamma}$, $\gamma=0.5772...$ is the Euler constant) and the change of variables,
$k \rightarrow q(x - {1\over2})$, $k' \rightarrow q(x' - {1\over2})$ to rewrite the
integral equation (\ref{schrodinger}) in the dimensionless form,
\begin{equation}
\int_{-1/2}^{+1/2} {dx' \over 2 \pi} \log(x-x')^2 \psi(x') + \phi(x) \psi(x) = 
\tilde\epsilon \psi(x)
\end{equation}
where 
\begin{equation}
\phi(x) = - {x + {1 \over 2} \over 2 \pi} \log({1 \over 2} + x)^2
+ {x - {1 \over 2} \over 2 \pi} \log({1 \over 2} - x)^2
\end{equation}
and the dimensionless eigenvalue $\tilde\epsilon$ is related to the dimensionful 
eigenvalue by 
\begin{equation}
\tilde\epsilon = {\epsilon(q) \over e^2 q} - {1\over 2 \pi} \log({C\over q^2}) \ .
\end{equation} 
The scaled equation is solved by the choice $\psi_o(x) \equiv 1$ which, being
nodeless, is associated with the lowest eigenvalue by the Perron-Frobenius theorem.
The corresponding eigenvalue is $\tilde\epsilon_o = {-1/\pi}$, whence we obtain
the EMP dispersion,
\begin{equation}
\epsilon(q) = {e^2 q \over 2 \pi} \log({8\over e^\gamma q^2}) \ .
\end{equation}

We now turn to the $\Delta S^z=-1$ sector, which will yield a new edge excitation,
the edge spin wave (ESW), in addition to the bulk spin wave continuum found 
previously. The relevant integral equation is now (\ref{schrodinger}) with the simplified
notation, 
\begin{eqnarray}
T_{\downarrow q} (k-k') & \rightarrow & T_q (k-k') \nonumber \\
V_{\downarrow q} (k) & \rightarrow & V(k) = - \int_{-\infty}^0 
{dk' \over 2 \pi}T_0(k-k') \ .
\end{eqnarray}
(We have ignored the Zeeman energy which can be trivially added back to the various
dispersions we will compute below.) Note that for the ideal edge, the potential energy is independent of $q$ exactly as in the bulk case.
Figure 3a shows a plot of $V(k)$ for this case which exhibits a minimum at the
edge, as expected from the Hartree-Fock energy levels. This should lead us to
expect one or more spin wave states bound to the edge. However, there is a 
competing effect --- the kinetic energy is larger near the edge as well, for 
the ``particle'' can only hop to the left.
\begin{figure}[htbp]
  \begin{center}
    \leavevmode
    \epsfxsize=8cm
    \epsfbox{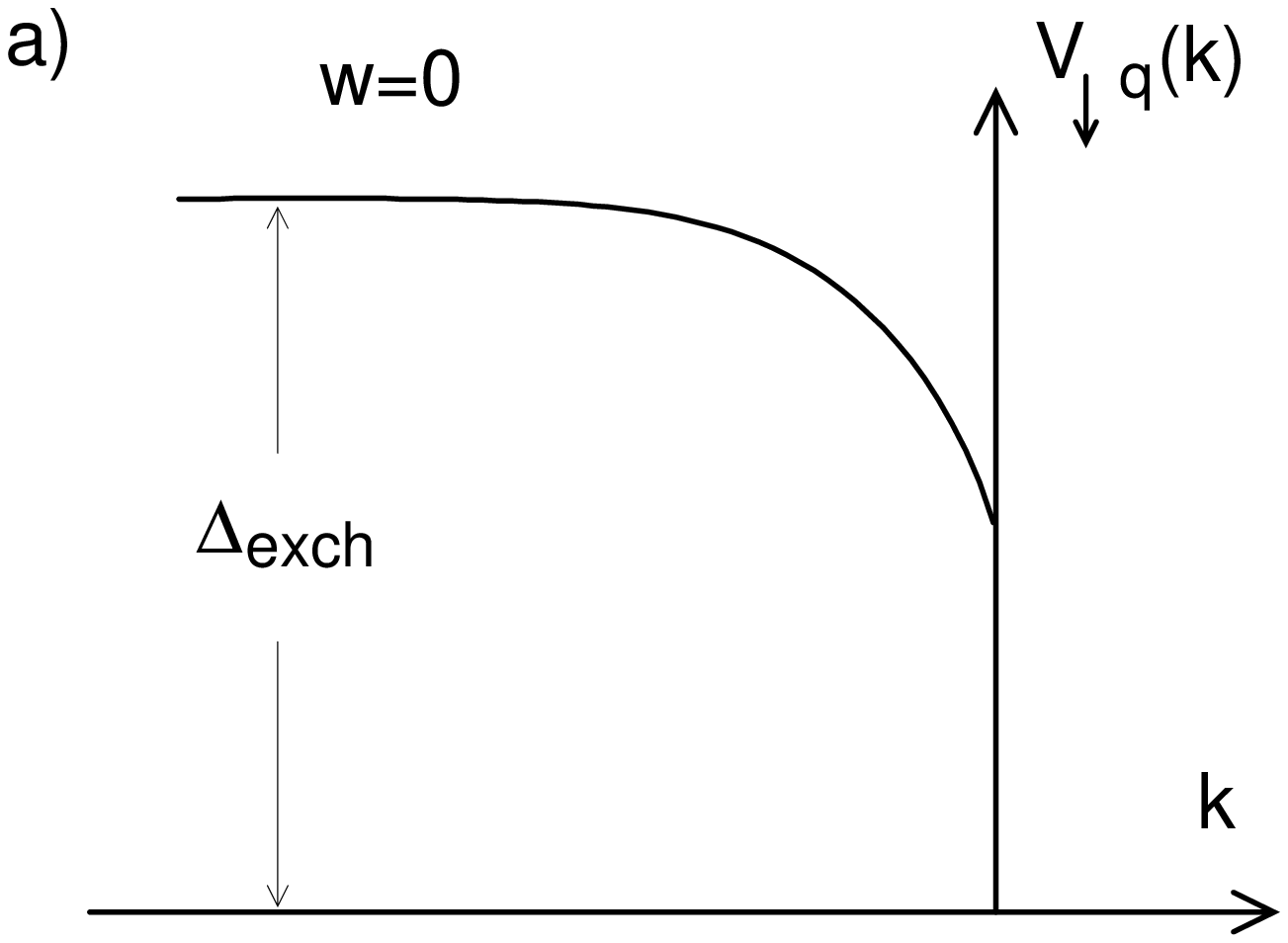}
    \epsfxsize=8cm
    \epsfbox{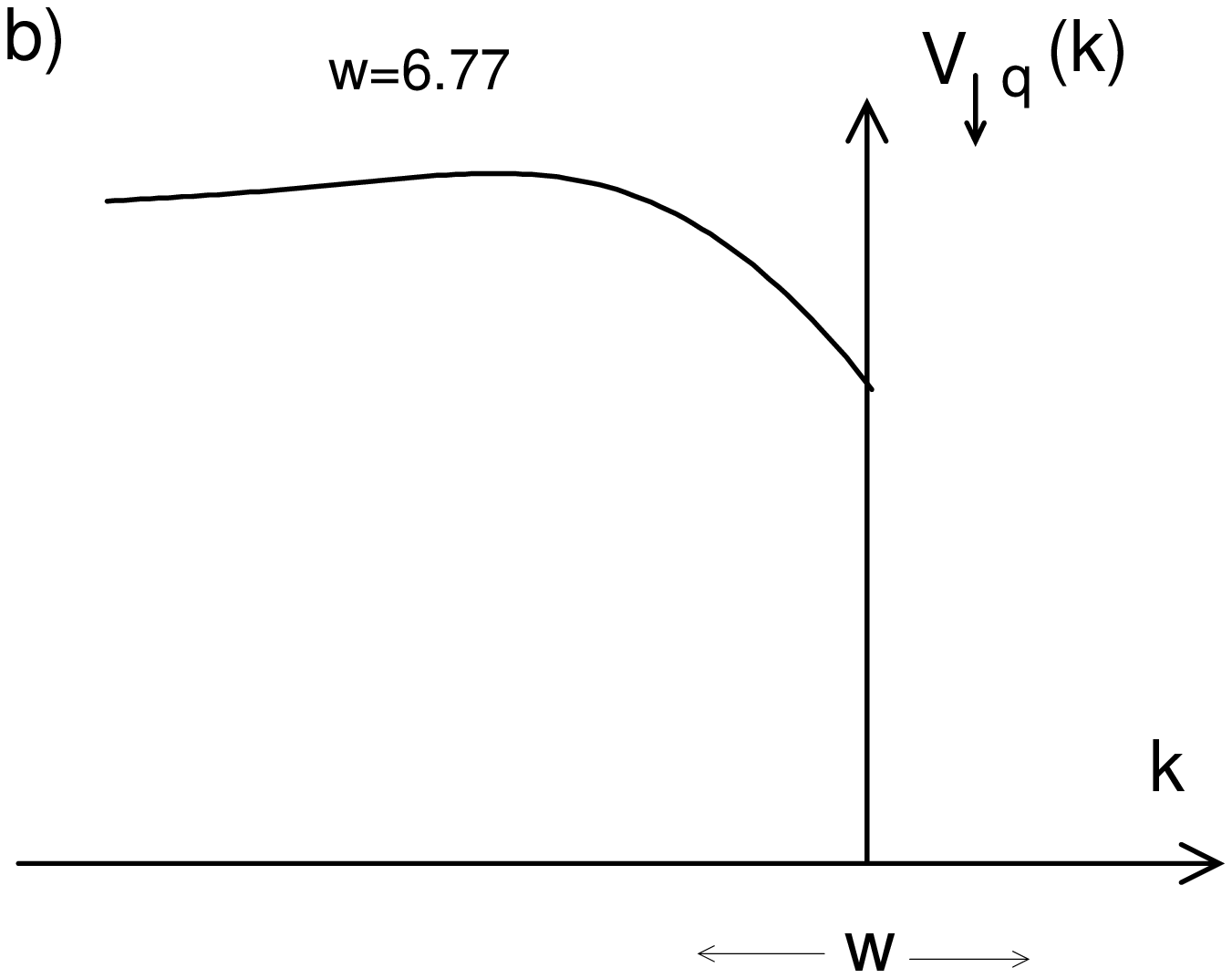}
  \end{center}
\caption{Potential energy for spin flip particle hole pairs for (a) the ideal edge 
and (b) a linear background with $w=6.77$ and $q=0.817$ for illustration.}
\label{fig3}
\end{figure}

Exactly at $q=0$, these effects compensate each other perfectly: the ground state
in this sector is given by the eigenfunction $\psi(k) \equiv 1$ with eigenvalue
zero. The eigenvalue can be checked by direct substitution and the Perron-Frobenius
theorem again shows that it is the ground state. (The existence of this state is 
not really a surprise --- it is just the statement that even for a geometry with
an edge, $S_{\rm total}^{-}$ gives rise to a degenerate state when acting on the 
ferromagnetic ground state.) It is important to note though, that $\psi(k) =1$ is 
really a {\it marginally bound} state as in the absence of the
attraction, the wavefunction would be suppressed at the edge. 

Consequently, at any finite $q$, where the matrix elements of the kinetic energy
are reduced by terms of $O(q^2)$, while the potential energy is unchanged, the
lowest energy state is indeed bound to the edge --- albeit with a localization
length that diverges as $q \rightarrow 0$. Although this is of academic interest,
we note that $T_q(k-k') \rightarrow 0$ as $q \rightarrow \infty$ and in that 
limit an arbitrarily large number of states become bound to the edge. Figure 4a
shows the first few states for a finite system with two edges, which clearly
exhibits a state bound to each edge as well as the first few continuum states.
The existence of a bound state {\it per edge} is a non-generic feature of the 
ideal edge problem which is not sensitive to the sign of $q$, i.e. whether
the particle is moved inwards or outwards.  Generically,
as we shall see, they exist for only one sign of $q$ and are therefore chiral.
Figure 4b illustrates the presence of multiple bound states at larger values
of $q$.
\begin{figure}[htbp]
  \begin{center}
    \leavevmode
    \epsfxsize=6cm
    \epsfbox{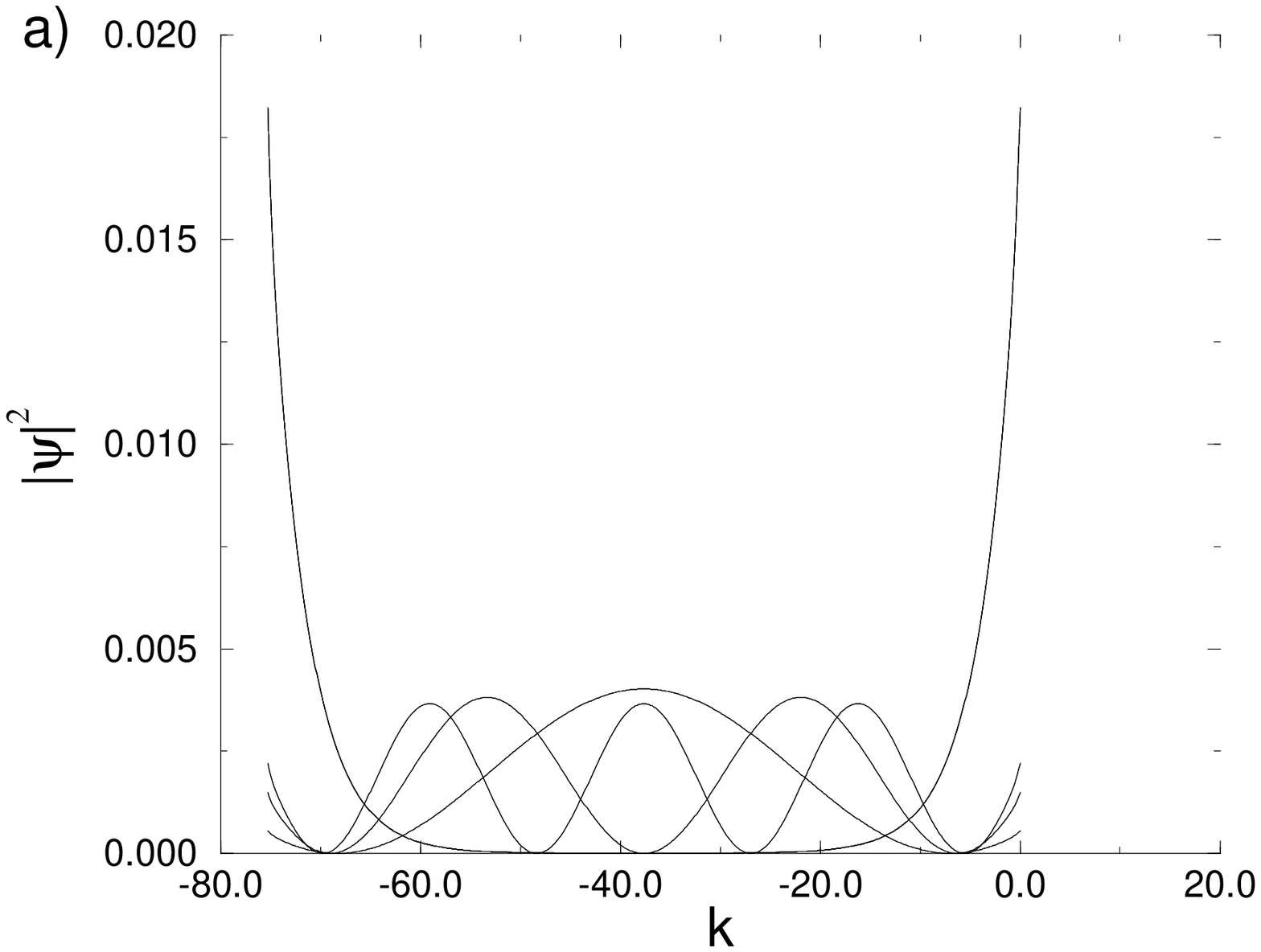}
    \epsfxsize=6cm
    \epsfbox{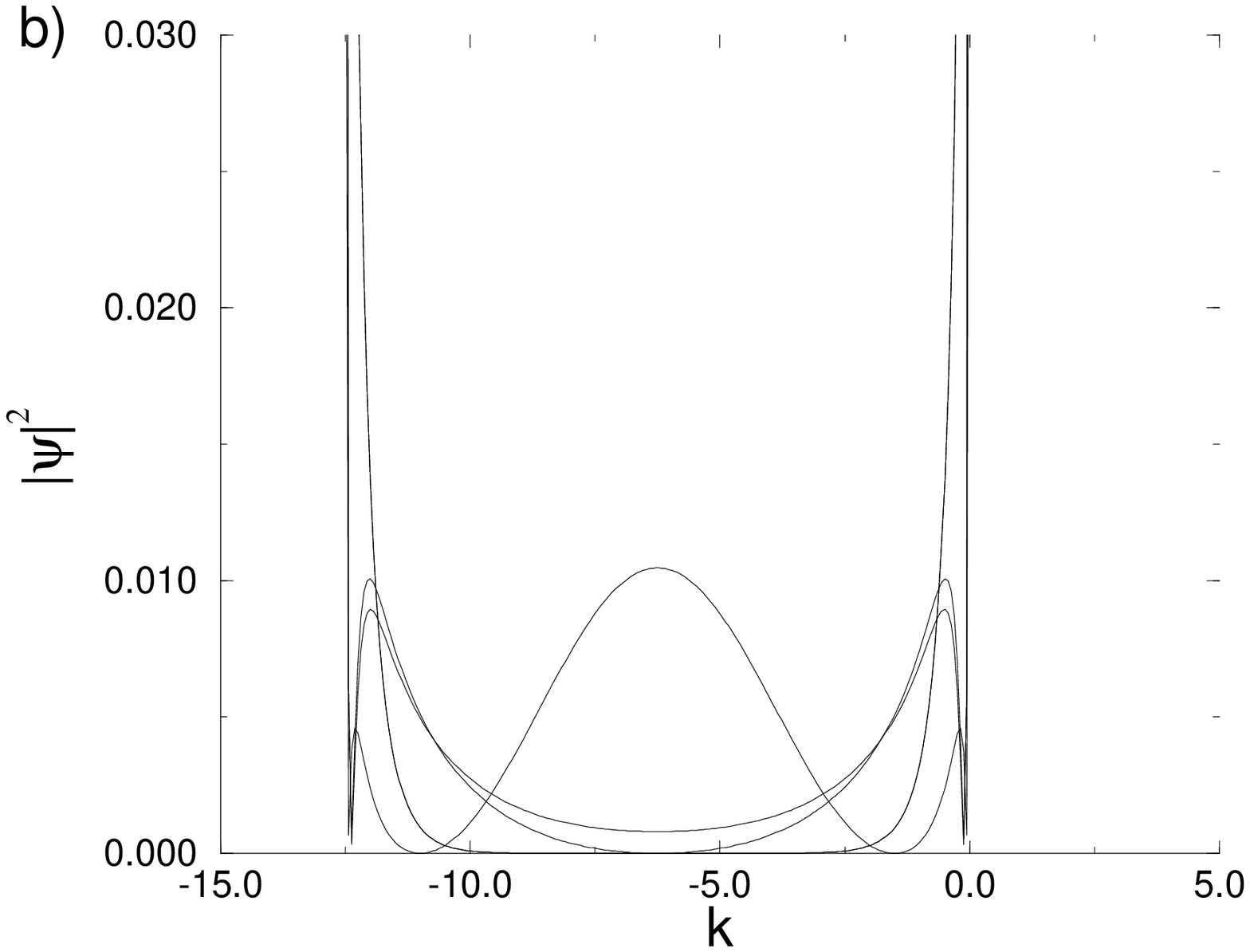}
  \end{center}
\caption{Probability densities for the lowest energy spin wave states in a 
system with two (ideal) edges. In (a), $q=0.628$ and one state is bound to each edge 
and in (b) $q=1.885$ and two states are bound to each edge. (For numerical reasons, 
data are shown for Hall bars of different width in (a) and (b).)}
\label{fig4}
\end{figure}

At small $q$, the localization length \cite{fn-length} of the ESWs is parametrically 
longer than the edge region which is $O(1)$ in our units. This enables us to get
a variational estimate of their properties at small $q$ by using the simple
exponential form $\psi(k) = \sqrt{2 \lambda}e^{ \lambda k}$. Far from the 
edge, this form is dictated by the recovery of translation invariance and it
has the feature that it correctly reproduces the exact answer at $q=0$ in the
limit $\lambda \rightarrow 0$. It is possible that this procedure gives 
asymptotically exact answers at small $q$ for some choices of interaction, 
but we do not have a proof of this. We expect to discuss this question 
elsewhere \cite{fn-wip}.

An equivalent procedure, which offers a different perspective, is to trade the 
integral equation for an approximate differential equation by the gradient 
expansion,
\begin{equation}
\psi(k') = \psi(k) + \psi'(k) (k'-k) + {1 \over 2} \psi''(k) (k'-k)^2 + \cdots
\end{equation}
Upon inserting in (\ref{schrodinger}) and integrating, we find, to $O(\psi'')$,
\begin{equation}
-\alpha_q(k) \psi''(k) + \beta_q(k) \psi'(k) + \gamma_q(k) \psi(k) = 
\epsilon(k) \psi(k)
\end{equation}
where
\begin{eqnarray}
-2 \alpha_q(k) &=& \int_{-\infty}^0 {dk' \over 2\pi} (k'-k)^2 T_q(k-k') \nonumber \\
\beta_q(k) &=&  \int_{-\infty}^0 {dk' \over 2\pi} (k'-k) T_q(k-k') \nonumber \\
\gamma_q(k) &=& \int_{-\infty}^0 {dk' \over 2\pi} [ T_q(k-k') - T_0(k-k') ] \ .
\end{eqnarray}
Away from the edge, $\beta_q(k)$ vanishes rapidly while $\alpha_q(k)$
and $\gamma_q(k)$ tend to constants $\alpha_q, \gamma_q$, and we get an effective 
Schr\"odinger equation, valid for long wavelength bulk spin waves.
A plane wave solution, $\psi(k) \sim e^{i k_x k}$ then has energy 
$\epsilon = \gamma_q + \alpha_q k_x^2$ which is easily seen to be the expansion
of our earlier result (\ref{bulksw}) to $O(k_x^2)$. For the Coulomb interaction,
$\alpha_q={1\over4} \sqrt{\pi \over 2} + {O}(q^2)$ and 
$\gamma_q={1\over4} \sqrt{\pi \over 2} q^2 + {O}(q^4)$.

In the same region, a bound state will exhibit purely exponential decay,
$\psi(k) \sim e^{\lambda k}$, in this approximation, which fixes
$\epsilon = \gamma_q - \alpha_q \lambda^2$. To complete the solution we need
to solve,
\begin{equation}
[-\alpha_q(k) \psi''(k) + \alpha_q \lambda^2 \psi(k)] + \beta_q(k) \psi'(k)
+ [\gamma_q(k) - \gamma_q] \psi(k) = 0 \ .
\end{equation}
At small $q$, $\gamma_q(k) - \gamma_q \sim O(q^2)$. Assuming that $\lambda^2$,
$\psi''/\psi \ll q^2$ which will be clear {\it a posteriori}, we can
neglect the first term. Integrating the remaining terms,
\begin{equation}
\int_{-\infty}^0 dk \,\beta_q(k) \psi'(k) 
+ \int_{-\infty}^0 dk \, [\gamma_q(k) - \gamma_q] \psi(k) =0 \ .
\end{equation}
If the $q \rightarrow 0$ limit is smooth in that $\psi'(k)$ vanishes uniformly
in the edge region (which is really equivalent to the assumption of a purely
exponential form for the bound state when combined with our earlier neglect
of $\psi''(k)$) we find that
\begin{eqnarray}
{\psi' \over \psi}|_{\rm edge} &=& \int_{-\infty}^0 dk [\gamma_q - \gamma_q(k)] /
\int_{-\infty}^0 dk \beta_q(k) \nonumber \\
&=& c q^2 + O(q^4)
\end{eqnarray}
where $c= {2 \over 3} ({2 \over \pi})^{3/2} = 0.33863\ldots$ for the Coulomb 
interaction. From this derivation it is
clear that this will apply also to the continuum states at low energies
for which $\psi''$ can be ignored as well.

This boundary condition then fixes $\lambda = c q^2 + O(q^4)$ which is indeed
much smaller than $q$ and yields the bound state energy $\epsilon(q) =
\gamma_q - \alpha_0 c^2 q^4 + \cdots$ which shows that the bound state splits
off below the spin wave continuum at $O(q^4)$. This means that if we try to
take a formal continuum limit in which the magnetic length is taken to zero
while the spin wave stiffness is held fixed, the edge bound states disappear 
($c\sim \ell$).
Consequently the ESWs are a matter of microscopic detail and cannot arise
generally in a continuum/sigma-model treatment. We will revisit the boundary
condition question below.

\section{Soft Confinement: Modes and Edge Reconstruction}

We turn now to the effect of softening the confinement away from the limit
of the ideal edge; a parallel discussion can be given for hardening the
confinement. We will assume that the modified confinement does not alter
the compact edge. Instead we study its effect on the edge excitation
spectrum which will allow us to identify two instabilities towards
reconstruction of the edge.

To this end we consider adding an extra potential $V_p(k)$ that exerts an
outward force on the electrons near the edge. Previously, the Hartree
potential vanished, but now it equals $V_p(k)$ for both the up and down
spin bands. In Figure 5 we plot two examples of this. The first (5a) is
uniformly repulsive near the edge and is qualitatively similar to the
effect produced by moving the compensating charge out of the plane of the
electrons. The second (5b) is the potential of the linear confinement with
$w > 0$ which gives rise to a dipole {\it in} the plane at the edge.
Note the special feature that it leaves the HF eigenvalues exactly at
the edge unshifted, with the down spin eigenvalue degenerate with those
deep in the bulk.
\begin{figure}[htbp]
  \begin{center}
    \leavevmode
    \epsfxsize=8cm
    \epsfbox{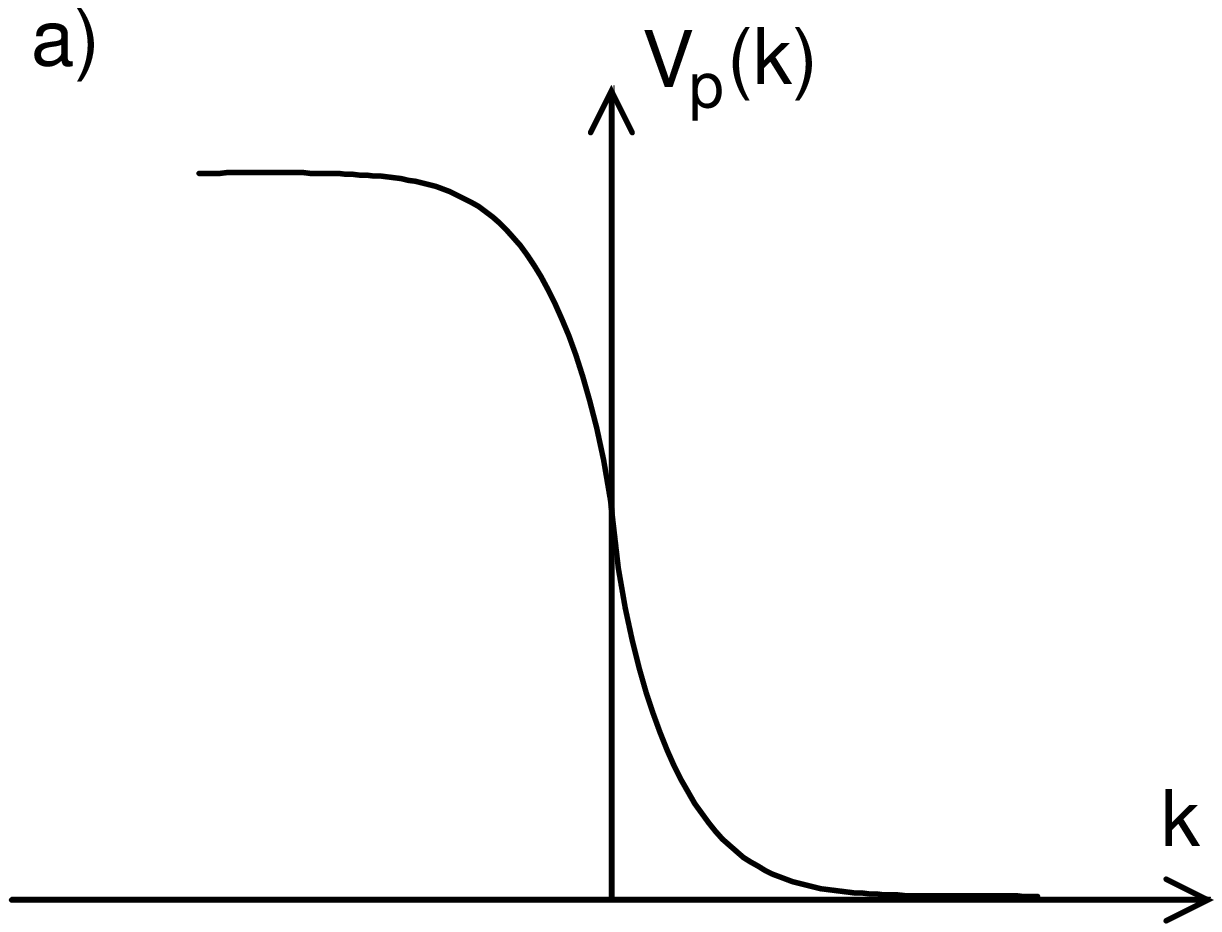}
    \epsfxsize=8cm
    \epsfbox{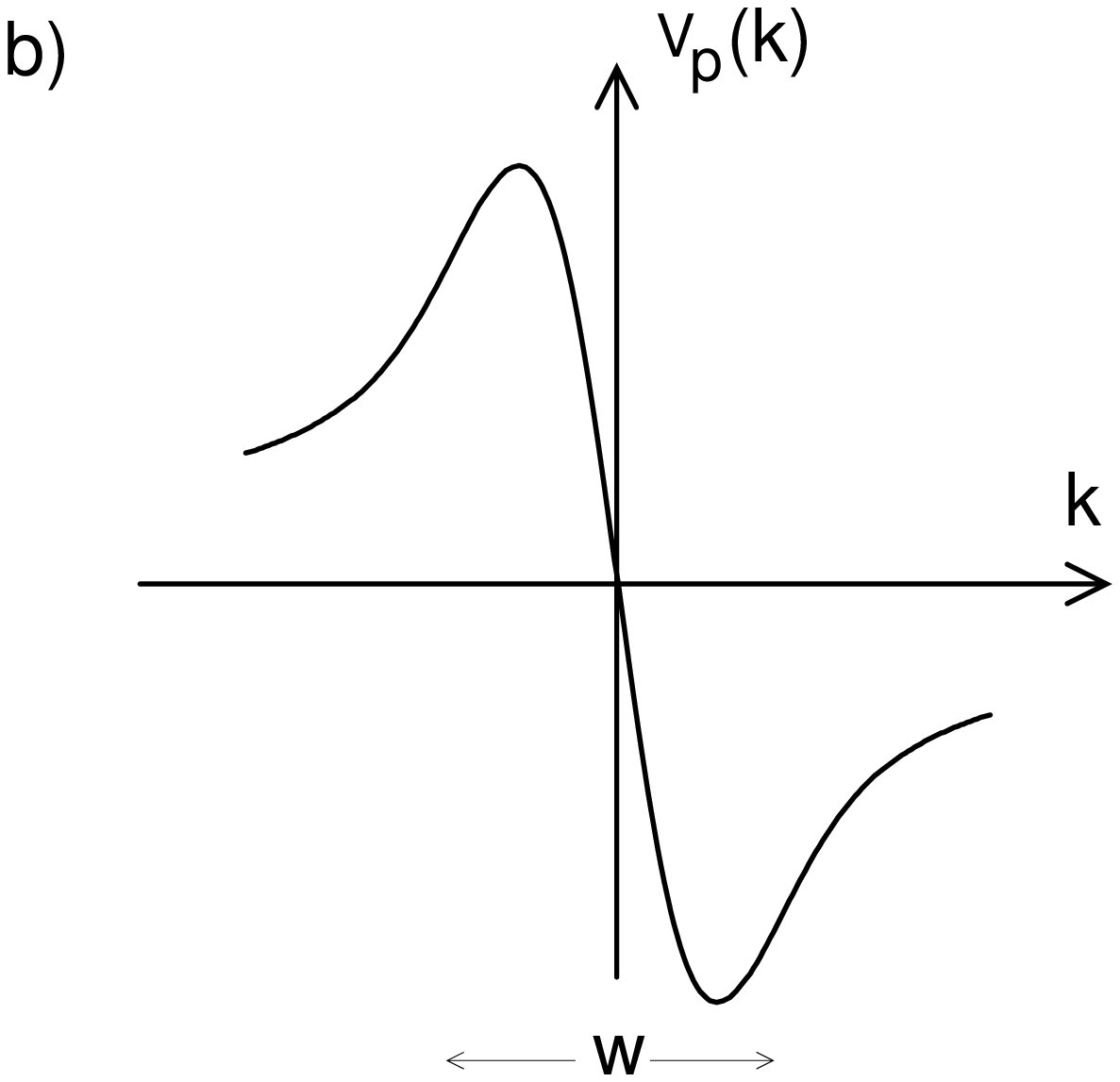}
  \end{center}
\caption{Perturbing edge potentials $V_p(k)$: (a) uniformly repulsive potential
due e.g. to background charge moved out of the plane of the electrons; 
(b) potential due to a linear in-plane background charge, $w=6.77$.}
\label{fig5}
\end{figure}

This modification has the effect of modifying the HF eigenvalues,
\begin{equation}
\delta \epsilon_{HF} (k \sigma) = V_p(k)
\end{equation}
and hence the potential energies for the particle-hole pairs by
\begin{equation}
\delta V_{\sigma q} (k) =  V_p(k+q) - V_p(k) \ .
\end{equation}
Consequently, the particle-hole pairs see an additional attraction near the edge
that varies with $q$ and typically is maximum for some $q > 0$.

By contrast, the $q<0$ pairs which exist in the $\Delta S_z=-1$ sector alone,
are repelled from the edge while the $q=0$ problem is left unaffected for 
both spin sectors. In Figure 6a we show examples of this effect for the potential
sketched in Figure 5a.
\begin{figure}[htbp]
  \begin{center}
    \leavevmode
    \epsfxsize=8cm
    \epsfbox{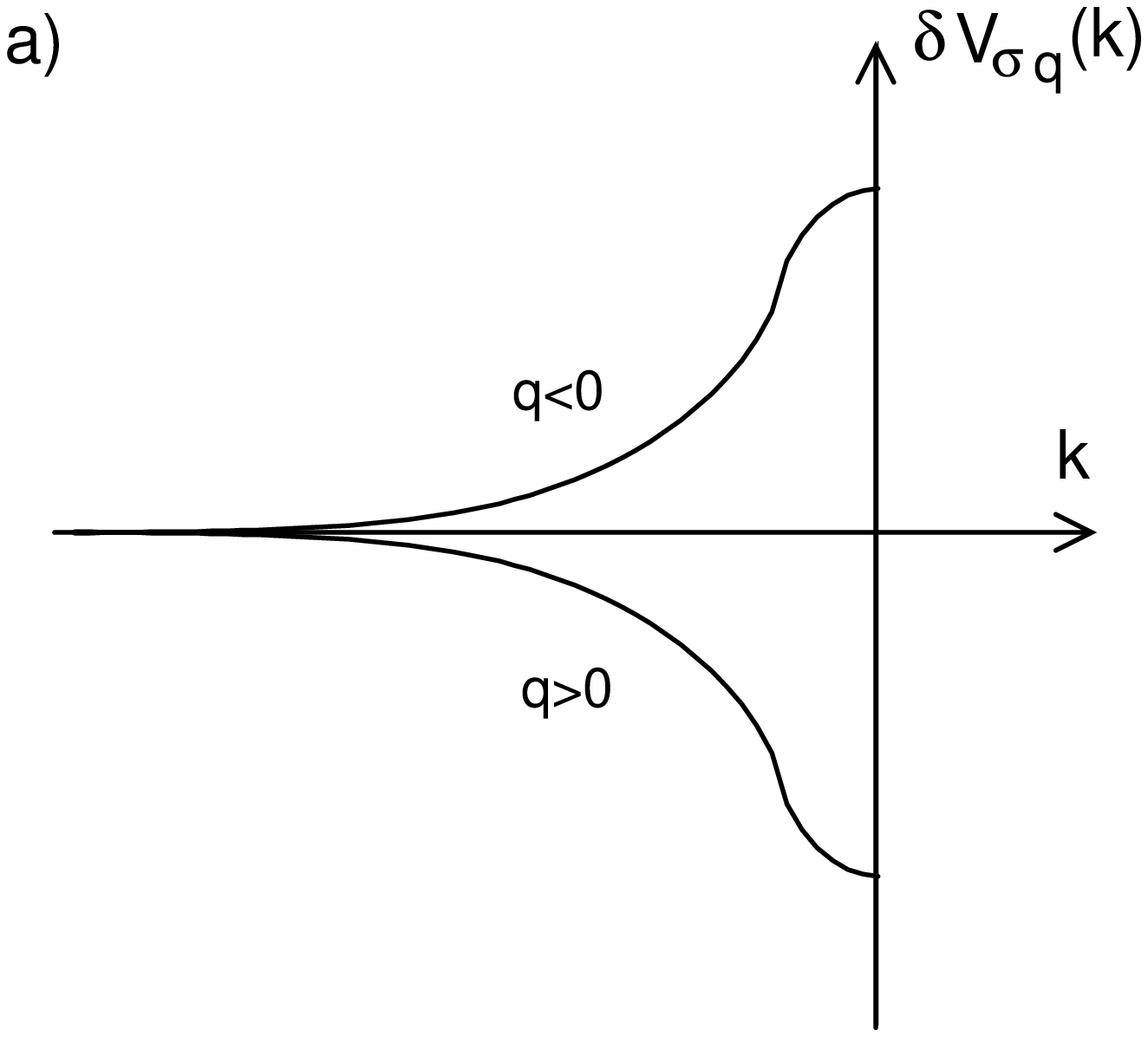}
    \epsfxsize=8cm
    \epsfbox{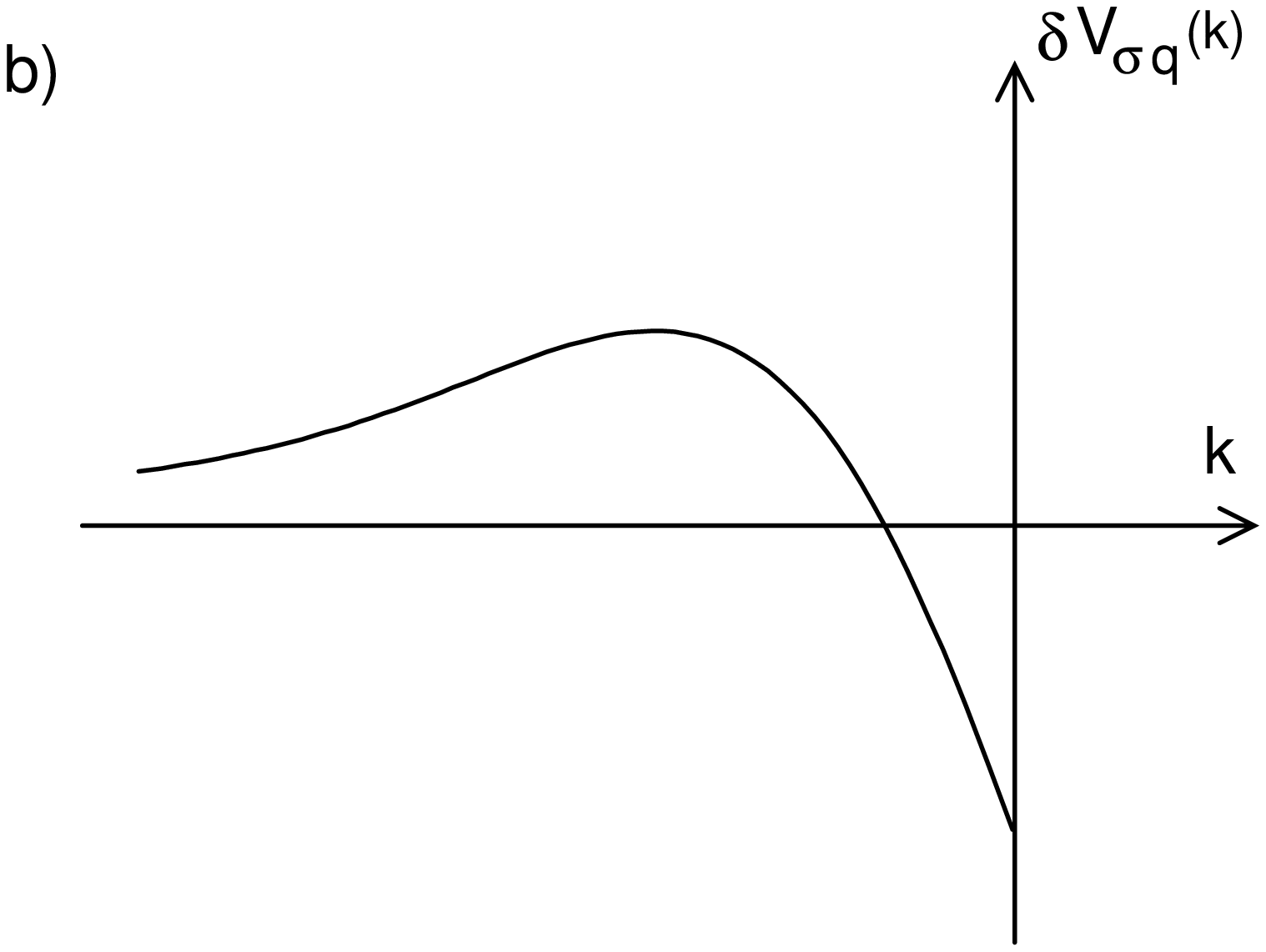}
  \end{center}
\caption{Shift in potentials $V_{\sigma q}$ due to the perturbing potentials in Figure
5a and 5b respectively, $|q|=0.817$.}
\label{fig6}
\end{figure}

The qualitative effect of this perturbation on the EMP spectrum is to localize
the mode wavefunctions $\psi(k)$ closer to the edge and to decrease the corresponding
eigenvalues. As the spectrum in a given $q$ sector is discrete, the latter effect 
can be reliably calculated perturbatively in $V_p(k)$. In the small $q$ limit
considered previously $\psi(k) = 1/\sqrt{q}$ for
$-q \le k \le 0$, whence,
\begin{eqnarray}
\Delta \epsilon_{\rm EMP}(q) & \sim &  \int_{-q}^0 dk \, |\psi(k)|^2 
\delta V_{\uparrow q}(k) \nonumber \\
& \sim & q V_p^{'}(0) 
\end{eqnarray}
to leading order in the perturbing potential. Note that this 
is just the energy gained by the particle-hole pair dipole in the edge
electric field. At larger $q$ and $V_p(k)$, the shift in the dispersion is a matter 
of more detailed computation and can lead to the development of a {\it minimum} 
(see Figure 7) that softens steadily as the confinement is weakened further.
\begin{figure}[htbp]
  \begin{center}
    \leavevmode
    \epsfxsize=12cm
    \epsfbox{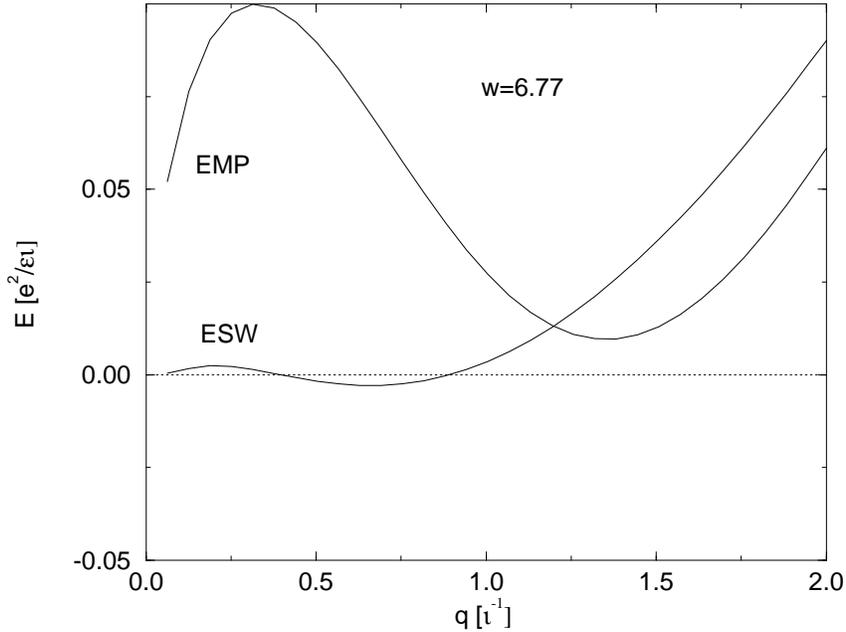}
  \end{center}
\caption{EMP and ESW dispersion relations for a linear confinement with $w=6.77$. Note
the softening of both modes at positive wavevectors and that the edge is already unstable
to textured reconstruction at small Zeeman energies.}
\label{fig7}
\end{figure}

If this new minimum dips below zero, the edge will reconstruct by macroscopically
occupying it, i.e. by a condensation of the corresponding bosons. If $q_0$ is
the wavevector at the minimum, then the corresponding boson creation operator
is,
\begin{equation}
B^\dagger(q_0) = \int_{-q}^0 dk \, \psi_{q_0}(k) b^\dagger(k,q_0) \ .
\end{equation}
This leads us to consider the state with $M$ condensed bosons,
\begin{equation}
[B^\dagger(q_0)]^M |G \rangle
\end{equation}
which is, in turn, recognized as the ``M boson" piece of the broken symmetry
state,
\begin{equation}
\Pi_{k \le 0} (u_k c^\dagger_{k\uparrow} + v_k c^\dagger_{k+q_0\uparrow} ) 
|0\rangle \ .
\label{cdwstate}
\end{equation}
with the identification, $\psi_{q_0}(k) = v(k)/u(k)$ close to the transition.
Evidently, (\ref{cdwstate}) describes a charge density wave state with a modulation
wavevector $q_0$ along the edge.

Returning to the $\Delta S_z =-1$ sector, we note that the perturbation has
two important effects. First, much as in the EMP case, it causes the
$q>0$ ESWs to soften and develop a minimum at a $q>0$. Second, generically,
it pushes the $q<0$ ESWs up into the bulk continuum where they cease to be 
bound states, although they may survive as resonances. In this fashion, the 
ESWs become chiral as well. (Note that for the case of a hardened confinement,
the ESWs will be chiral in the opposite sense to the EMPs.) 

As already remarked, the perturbing potential for the $\Delta S_z=-1$
particle-hole pairs vanishes at $q=0$ and hence the ground state 
wavefunction of that problem is still $\psi(k)=1$; indeed, this is
still a consequence of the ferromagnetic ground state being an eigenstate
of the squared total spin operator. Consequently, the bound states must again be
localized on a length scale that diverges at small $q$. For a generic
potential, we can again employ the variational choice $\psi(k)=
\sqrt{2 \lambda} e^{\lambda k}$ to obtain,
\begin{eqnarray}
\lambda &=& \sqrt{32 \over \pi} \, \Bigl[V_p(-\infty) - V_p(0) \Bigr]\, q \nonumber \\
\epsilon(q) &=& \gamma_q - \sqrt{32 \over \pi}\, \Bigl[V_p(-\infty) - V_p(0)
\Bigr]^2 \, q^2 \ .
\label{q2}
\end{eqnarray}
Evidently,  bound states exist only for $q>0$ and the ESWs are now chiral,
as advertised. Note that the localization length is much smaller, $O({1\over q})$
as against $O({1\over q^2})$ for the ideal edge, and that the dispersion is 
accordingly much softer as it now departs at $O(q^2)$ from that of bulk spin
waves.

It is clear from (\ref{q2}) that the case of the linear confinement is special, in that
$V_p(-\infty)=V_p(0)$. In this case the particle-hole pair potential (see Figure
6b) is both attractive and repulsive. At small $q$, $\delta V_{\downarrow 
q}(k) \sim q V_p'(k)$ and hence its integral $\int_{-\infty}^0 dk \,
\delta V_{\downarrow q}(k) = q[V_p(0) - V_p(-\infty)]$, vanishes 
to leading order in $q$. As a consequence, we obtain the modified variational
results,
\begin{eqnarray}
\lambda &=& \sqrt{2 \over \pi} \, \Bigl[{4 \over 3 \pi}  + 
2 |V_p'(0)| \Bigr] \, q^2 \nonumber \\
\epsilon(q) &=& \gamma_q -  \sqrt{2 \over \pi} \, \Bigl[{2\over 3 \pi} + |V_p'(0)| 
\Bigr]^2 \, q^4
\label{q4}
\end{eqnarray}
which differ from those for the ideal edge only in the increase in the coefficient $c$.
In particular, there are still ESWs of {\it both} chiralities at small $q$ whose
localization length is still $O({1\over q^2})$.

As with the EMPs, softening can lead to the development of an ESW minimum 
at a finite $q_0 > 0$ (see Figure 7), which can, again, dip below zero energy. If that happens, 
in complete analogy to our treatment of the EMP
softening, one can see that the edge reconstructs to a state of the form,
\begin{equation}
\Pi_{k \le 0} (u_k c^\dagger_{k\uparrow} + v_k c^\dagger_{k+q_0 \downarrow} )
|0\rangle
\end{equation}
which is the spin textured edge studied previously.

One important point emerges from (\ref{q4}). For a linear confinement, ESW 
softening can take place only at $O(q^4)$, which is a much less
efficient process than the generic softening at $O(q^2)$ detailed in (\ref{q2}). 
On tracking the relative softenings of the EMP and ESW minima, we find that the
phase diagram for the linear confinement is dominated by the charge density wave 
instability \cite{unpub}; this has also been noted by Franco and Brey from the completely
equivalent Hartree-Fock analysis of the reconstructed edge \cite{brey}. A generic 
confinement however, will lead to much softer ESWs and hence should lead
to a much stronger tendency to textured reconstruction. 

\section{More on Boundary Conditions}

In our analysis of the ideal edge spin wave problem, we noted that
our variational calculation is equivalent, at long wavelengths
perpendicular {\it and} parallel to the edge, to imposing the boundary
condition $\psi'(0)=cq^2 \psi(0)$ on the differential equation
$-\alpha_q \psi'' = (\epsilon -\gamma_q)\psi$. A similar reformulation is possible
for the non-ideal edge as well, where the bound states can be obtained
from the modified boundary conditions $\psi'(0)= \lambda \psi(0)$
with $\lambda$ given by equations (\ref{q2}) and (\ref{q4}). At least for 
the ideal edge, where we have studied the wavefunctions in some detail, the
boundary condition seems to work fairly well at moderately small
wavenumbers (there is a finite size problem that we have not overcome
at really small $q$) and hence we would like to explore this notion
a bit further. We should note that the existence of {\it some}
boundary condition reformulation of the problem is quite likely at
small $q$ where the bound state size must diverge. One can simply
rescale in units of this size and shrink the boundary region to
zero. If the boundary by itself does not induce subexponential
corrections to $\psi(k)$, it will end up giving rise to an effective
boundary condition.

The Landau gauge analysis identifies the state,
\begin{equation}
|\psi_q\rangle = \int {dk \over 2 \pi} \, \psi_q(k) \, c^\dagger_{k+q \downarrow}
c_{k \uparrow} |G\rangle
\end{equation}
with the {\it one-dimensional} spin wave eigenfunction $\psi_q(k)$.
At long wavelengths, this can be written in the suggestive form,
\begin{equation}
|\psi_q^{(2)} \rangle = \int dx \int dy \, e^{iqy} \psi_q(x) \,
\Psi^\dagger_{\downarrow}(x,y) \Psi_{\uparrow}(x,y) |G\rangle
\label{swstate}
\end{equation}
in terms of the {\it two-dimensional} spin wave eigenfunction
$e^{iqy} \psi_q(x)$. This indicates that in more general geometries,
we can obtain spin wave states by solving for the appropriate 
eigenfunctions of $-\alpha_q \nabla^2$ and identifying them with states
via (\ref{swstate}). It also suggests that we replace our boundary condition 
by the two-dimensional form $\partial_x \psi = -c \partial_y^2 \psi$
which can also be generalized, along with its analogs for non-ideal
edges, to other shapes for the edge. In physical units,
this is $\partial_x \psi = - c \ell \partial_y^2 \psi$ wherein the 
explicit $\ell$ leads to a ``trivial'', Neumann boundary condition
in the continuum limit. For the generic non-ideal edge however,
we get, according to (\ref{q2}), $\partial_x \psi \propto \partial_y\psi$ 
which does survive the continuum limit. 

This prescription for constructing spin wave states enables us to make 
contact with the interesting work of Oaknin et al \cite{oaknin}, who 
independently arrived at it in their study of spin waves on a disc with 
one important difference---reasoning directly in the continuum, they concluded 
that a Neumann
boundary condition was appropriate regardless of the details of
the confinement. We have seen that this is not ``sufficiently'' 
correct. The boundary conditions needed to reproduce the ESWs
are different, although they {\it do} approach the Neumann condition
for asymptotically long wavelengths along the edge. This is clear
in the semi-infinite geometry that we have studied in the bulk of
this paper but it is less clear in the case of the disc. Here
Oaknin et al obtained extremely impressive overlaps between the
lowest energy exact eigenstates for a system of $N=30$ electrons
and spin waves with $\Delta $ (the net angular momentum of the
spin flip pair which replaces $q$ in our geometry) ranging from
1 to 30, and the states generated by their prescription.

To check that our boundary conditions are more appropriate than
theirs even for a disc, we have carried out a similar numerical
test, also for a system of 30 electrons. To simplify the computations,
we have used a hard core potential $V(\bf x) = \delta^2(\bf x)$
and the disc analog of our ideal edge with only the exchange 
potential entering the Hartree-Fock description. By hypothesis,
this should not matter to Oaknin et al. We 
extract the value $c=1/\sqrt{2 \pi}$ from an analysis on the
semi-infinite plane and then postulate the boundary condition,
\begin{equation}
{\partial \psi \over \partial r}\Bigr|_{r=R} = - {c\over R^2} 
{\partial^2  \psi \over \partial \theta^2} 
\label{discbc}
\end{equation}
at the radius $R$ of the disc. The eigenfunctions of the Laplacian
on the disc are
\begin{equation}
\psi_{k\Delta}(r,\theta) = J_\Delta(kr) \, e^{i\Delta\theta}
\end{equation}
which then must satisfy $kJ'_\Delta(k_\Delta R)= c {\Delta^2 \over R^2} 
J_\Delta(k_\Delta R)$ .
(The other Bessel function, $I_\Delta(kr)$, leads to new solutions only at
large values of $\Delta$, $\Delta > R/c$, which are beyond the validity
of the long wavelength approximation and so we do not consider it here.)
On comparing the exact eigenstates in the single spin flip sector with
our ansatz (first entry) as well as the Neumann ansatz (second entry), we 
find that the overlaps for the ground state wavefunction range from 
(0.999901, 0.99981) for $\Delta =1$, via (0.997695, 0.99436) for $\Delta =3$,
to (0.905964, 0.734507) for $\Delta =10$. Evidently our ansatz does better, 
but the improvement is barely perceptible at small $\Delta$ where one has
reason to trust this procedure! This is due partly to the smallness of the right hand
term in (\ref{discbc}) which causes it to mimic the Neumann condition for
the small $\delta$ values where a continuum description is appropriate, and
partly to the nodeless character of the ground state wavefunction which 
makes all overlaps large. 
In sum, while it is possible to detect the requirement of
a non-trivial boundary condition for a small system such as the one considered
by Oaknin et al, the effect is quite small and was naturally overlooked by them. 
As the system size is increased, the effect will increase, for in the infinite 
radius limit we approach the problem of the linear edge---and in that case, the 
lowest energy eigenfunctions with the Neumann condition which are not bound to 
the edge, have zero overlap with the true lowest energy eigenfunctions. Our 
boundary condition, by contrast, is designed to reproduce exactly those states.

A point of explanation is on order. Readers may be puzzled by our unwillingness to
make more of the substantial discrepancy in overlaps at $\Delta =10$ and other
readers familiar with the Oaknin et al paper might be puzzled that the overlap
they report for the same value of $\Delta$ is much larger. In reporting overlaps
we have literally used the prescription embodied in (\ref{swstate}). This requires
evaluating an integral involving two Landau orbitals and the Bessel function over
the entire real line. This becomes meaningless at large $\Delta$ once one of the 
Landau orbitals is well outside the disc for at least some locations of the
spin-flip pair. Indeed this causes the Neumann states to develop nodes and their
overlap with the true ground state to plummet. Our boundary condition does better
as it pushes the nodes of the Bessel functions further out, but clearly the whole 
prescription is no longer well motivated in this limit. The results reported by
Oaknin et al are {\it not} obtained by the literal application of (\ref{swstate});
instead they are obtained by a procedure that is equivalent to approximating the 
Bessel functions by polynomials which no longer exactly obey the Neumann condition
at the boundary but are then able to avoid the sign problem coming from integrating 
outside the disc \cite{tejedor-pc}. While it is interesting that their final prescription 
does as well as it does, it is still a somewhat {\it ad hoc} fix for the problems
of the formalism at large $\Delta$ which does not appear susceptible to generalization 
to other geometries, e.g. the semi-infinite plane. Consequently we will not pursue
it further in this paper.

Finally, we note that the low energy spin physics of the $\nu=1$
state is expected to be governed by the ferromagnetic $O(3)$
sigma model appropriate to all isotropic ferromagnets, e.g. the
magnetization of the $\nu=1$ state at finite temperatures has
been analyzed in this fashion by Read and Sachdev \cite{rs}. For the 
ideal edge, the boundary condition on the spin flip pair wavefunction
translates into the boundary condition,
\begin{equation}
{\bf n} \cdot ( \partial_x {\bf n} \times c {\partial^2_y {\bf n}})
=0
\label{sigmabc}
\end{equation}
where ${\bf n}(x)$ is the vector order parameter field. Indeed one can 
find, in this particular case, a similar phenomenon in a lattice ferromagnet in
which the bonds between the spins on the edge are weakened 
from their bulk values. If the bonds along the edge have strength
$J'$ while those in the bulk have strength $J$, there is a set
of edge spin waves with localization length $\lambda = (J-J')a q^2/
J $ at momentum $q$ parallel to the edge. In this
case one can arrive at the boundary condition (\ref{sigmabc}) by directly
considering the equations of motion for the edge spins. This
boundary condition, and all the others we have considered in
this paper, have the feature that they lead to currents even
at the boundary which would be unnatural in a purely continuum
approach. We emphasize that allowing for these microscopic
currents (which represent the motion of the spins at the edge)
is necessary in order to capture the edge spin waves within
the continuum formulation, the latter in turn are interesting
for their role in textured reconstruction.

\section{Concluding Remarks}

The foregoing analysis has been confined to the TDHF approximation and we have
only kept track of single particle hole pair states. The states
in the $\Delta S_z=0$ sector will only mix among themselves and
the resulting interacting problem of many pairs is still described
by the theory of a single non-interacting chiral boson at 
low energies, this is the simplest instance of Wen's bosonic
description of quantum Hall edges \cite{stone}. The dispersion of the bosonic
mode is given {\it exactly} by our calculation at long wavelengths.
The $\Delta S_z=-1$ sector will involve
a mixing between the ESWs and the EMPs which will renormalize the
ESW dispersion. However it is easy to see that energy and
momentum conservation rule out decay of the ESWs by EMP emission
at low momenta (for $q<\hbar v/\rho_s$ where $v$ is the EMP velocity
and $\rho_s$ is the ESW stiffness, e.g. given by our expressions (\ref{q2})
and (\ref{q4})) and they continue to be well defined excitations in that region.
Our arguments are less robust at the larger wavevectors that are
more germane to actual reconstruction instabilities and while we
feel quite confident that the qualitative physics uncovered in this
paper is equally applicable in that region, we do not have a sense
of the quantitative error involved in restricting calculations to the
TDHF approximation.

To summarize: We have shown that the edge dynamics of a polarized compact
$\nu=1$ state exhibits two modes, the EMP and an ESW. The existence of the 
latter is a new result, and this mode has several unusual properties, most notably
its lack of a consistent chirality for all confinements. Both modes can soften 
with softening confinement and cause the edge to reconstruct by charge density 
wave or spin texture formation, respectively. We expect that this dual physics 
generalizes {\it mutatis mutandis} to other quantum Hall ferromagnets.

We expect that the most likely experimental detection of the physics described in
this paper is through the observation of the resulting edge reconstructions and
their sensitivity to various edge parameters, e.g. the Zeeman energy in the case
of textured reconstruction. Direct detection of the ESW for a compact edge appears
to be a challenge.

\acknowledgements
We are grateful to S. M. Girvin, T. H. Hansson, S. A. Kivelson. J.-M. Leinaas 
and A. H. MacDonald for enlightening comments and to R. Moessner for a
careful reading of the manuscript.
This work was supported in part by NSF grant No. DMR-96-32690, the A. P. Sloan 
Foundation and the Institute for Advanced Study (SLS), and the Swedish Natural 
Science Research Council (AK).


\begin{references}
\bibitem{fn-eps} The background dielectric constant of the host semiconductor
can be included by replacing $e^2$ by $e^2/\epsilon$ in this and other 
expressions in this paper.
\bibitem{fn-length} As we are discussing wavefunctions defined on the set of
momenta, strictly speaking, the localization ``length'' has units of a momentum.
However it is clear from the structure of the lowest Landau level wavefunctions,
that this is the same as a physical length upon multiplication by the square
of the magnetic length. This is invisible in our notation, where $\ell=1$ and hence
we shall not make the distinction in what follows.
\end{references}
\end{document}